\documentclass[aps,prl,twocolumn,amsmath,amssymb,superscriptaddress,floatfix]{revtex4-1}

\usepackage{graphicx}
\usepackage{multirow}
\usepackage{color}
\usepackage{bm}

\renewcommand{\Vec}[1]{\mbox{\boldmath$#1$}}

\def\t#1{\textrm{#1}}

\def\braket#1{\langle #1 \rangle}
\def\n{\nonumber \\ }

\begin{document}

\title{
Charge and spin transport in edge channels of 
a $\nu=0$ quantum Hall system on the
surface of topological insulators
}

\author{Takahiro Morimoto}
\affiliation{Condensed Matter Theory Laboratory, 
RIKEN, Wako, Saitama, 351-0198, Japan}
\author{Akira Furusaki}
\affiliation{Condensed Matter Theory Laboratory, 
RIKEN, Wako, Saitama, 351-0198, Japan}
\affiliation{RIKEN Center for Emergent Matter Science 
(CEMS), Wako, Saitama, 351-0198, Japan}
\author{Naoto Nagaosa}
\affiliation{RIKEN Center for Emergent Matter Science 
(CEMS), Wako, Saitama, 351-0198, Japan}
\affiliation{Department of Applied Physics, The University of 
Tokyo, Tokyo, 113-8656, Japan}

\date{\today}

\begin{abstract}
Three-dimensional topological insulators of finite thickness
can show the quantum Hall effect (QHE) at the filling factor $\nu=0$
under an external magnetic field if there is a finite potential
difference between the top and bottom surfaces. 
We calculate energy spectra of surface Weyl fermions in the $\nu=0$ QHE
and find that gapped edge states with helical spin structure
are formed from Weyl fermions on the side surfaces under certain conditions.
These edge channels account for the nonlocal charge transport in the $\nu=0$ QHE which is observed in a recent experiment on (Bi$_{1-x}$Sb$_x$)$_2$Te$_3$ films.
The edge channels also support spin transport due to the spin-momentum locking.
We propose an experimental setup to observe various spintronics functions such as spin transport and spin conversion.
\end{abstract}

\maketitle

\textit{Introduction ---}
The quantum Hall effect (QHE) is a representative topological 
quantum phenomenon where edge channels 
play an essential role in low-energy transport \cite{QHE1,QHE2}.
The emergence of gapless edge channels is closely tied to
nontrivial topology of
gapped electronic states in the bulk, which is
the property called bulk-edge correspondence.
For topological insulators (TIs) \cite{TI1,TI2},
the bulk-edge correspondence dictates that
any surface of a three-dimensional (3D) TI with a nontrivial
$\mathbb{Z}_2$ index has a single (or an odd number of) flavor(s) of
Weyl fermions with spin-momentum locking.
The surface Weyl fermions are predicted to give
a variety of novel phenomena such as the quantized 
topological magneto-electric (ME) effect \cite{TME} 
and monopole-like magnetic field distribution induced by
a point charge \cite{Monopole}. 
Another remarkable phenomenon is
the quantized anomalous Hall effect
{\it without} an external magnetic 
field \cite{Haldane, Onoda, Fang, Xue}.
By contrast,
in an external magnetic field,
a single flavor of
Weyl fermions are expected to show a ``half-integer'' QHE with the Hall conductance
$\sigma_{xy}= (n + \frac{1}{2} ) e^2/h$ ($n\in\mathbb{Z}$),
which has a $\frac12$-shift due to their nontrivial Berry phase,
compared with the quantized Hall conductance of
non-relativistic electrons,
$\sigma_{xy}= n e^2/h$ ($n\in\mathbb{N}$).

However, the ``half-integer'' QHE is forbidden by the
Nielsen-Ninomiya theorem \cite{Ninomiya},
which dictates that there should be an even number of flavors of
Weyl fermions in the first Brillouin zone,
implying that the half-integer QHE is not observable in reality.
For example, Weyl fermions in graphene have four flavors
from spin and valley degrees of freedom, yielding
$\sigma_{xy}= 4 (n + \frac{1}{2} ) e^2/h$ \cite{graphene}.
As for a 3D (strong) TI, excitations at all surfaces 
have to be considered together.
In a magnetic field applied perpendicular to
the top and bottom surfaces of a 3D TI, Weyl fermions on the both
surfaces contribute to $\sigma_{xy}$, yielding
$\nu:=\sigma_{xy}/(e^2/h)= n_T + n_B +1$ 
with $n_T+\frac12$ and $n_B+\frac12$ coming from the top and bottom surfaces,
respectively.
The integers $n_T$ and $n_B$ are LL indices of the highest occupied LLs
at the top and bottom surfaces.
When the same number of LLs are filled at the top and bottom surfaces
($n_T=n_B$),
the Hall conductivity is quantized at an odd integer \cite{DHLee}.
However, the QHE of surface Weyl fermions
with Hall plateaus at $\nu=2, 3, 4,\ldots$ is experimentally
observed in 3D HgTe \cite{Brune}.
More recent experiments on (Bi$_{1-x}$Sb$_x$)$_2$Te$_3$ films
have reported plateaus at $\nu=0$ and $\pm1$ as the gate voltage
(Fermi energy) is changed \cite{Yoshimi,Xu}.
The appearance of $\nu=0$ plateau violating the odd-integer rule 
indicates that the degeneracy between the top and bottom surfaces
is lifted in (Bi$_{1-x}$Sb$_x$)$_2$Te$_3$ films \cite{Yoshimi,Xu}.
Moreover, a transport measurement on (Bi$_{1-x}$Sb$_x$)$_2$Te$_3$ films revealed the existence of nonlocal transport in the $\nu=0$ QHE \cite{PrivateCommun}.
The nonlocal transport appears to conflict with the $\nu=0$ QHE because of the absence of the chiral edge mode supporting the nonlocal transport.

In this paper, we study the $\nu=0$ QHE
(i.e., $n_T+n_B+1=0$) of Weyl fermions
on the surfaces of a 3D TI of finite thickness. 
Compared with the $\nu=0$ QHE in graphene,
which has been studied actively
both experimentally \cite{Abanin,Joe, Giesbers, Zhao}
and theoretically \cite{Lee,Sarma},
the QHE in 3D TIs has the following unique features.
First, the Weyl fermions on the TI surfaces are two-component spinors of real
spin-$\frac12$ degrees of freedom (as opposed to pseudo-spins in graphene),
and the edge channels support both charge and spin transport.
Second, in a magnetic field the Weyl fermions on the top and bottom surfaces
(which are perpendicular to the field) form Landau levels (LLs),
while those on the side surface (parallel to the magnetic field) are
not directly affected by the field.
However, one cannot consider independent 2D top and bottom surfaces of
$\sigma_{xy}=\pm e^2/2h$ with edge channels.
The side surface which connects the top and bottom surfaces must be
taken into account when discussing the edge channels.
The energy spectrum of the Weyl fermions on the side surface is
discretized
in a thin 3D TI, yielding well-defined edge channels
under the condition of Eq.~(\ref{eq: condition for edge channels}).
These edge channels account for the nonlocal charge transport in the $\nu=0$ QHE observed in the recent experiment \cite{PrivateCommun}.
Furthermore, since these edge channels are spin-momentum locked,
the $\nu=0$ QHE of TI thin films offers a new arena for various spintronics functions such as spin transport and spin conversion.

\begin{figure}[tb]
\begin{center}
\includegraphics[width=0.8\linewidth]{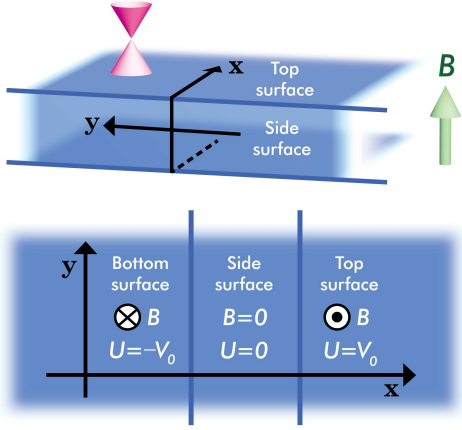}
\end{center}
\caption{
Schematic picture of the two-dimensional (2D) model for the surface Weyl
fermions of a topological insulator in a magnetic field.
A 2D plane in the lower panel represents the top, bottom and side surfaces,
where the strength of the magnetic field is $\pm B$ for $x>d$ and $x<-d$
and vanishing for $-d<x<d$.
}
\label{Fig: surface of TI}
\end{figure}

\begin{figure}[tb]
\begin{center}
\includegraphics[width=\linewidth]{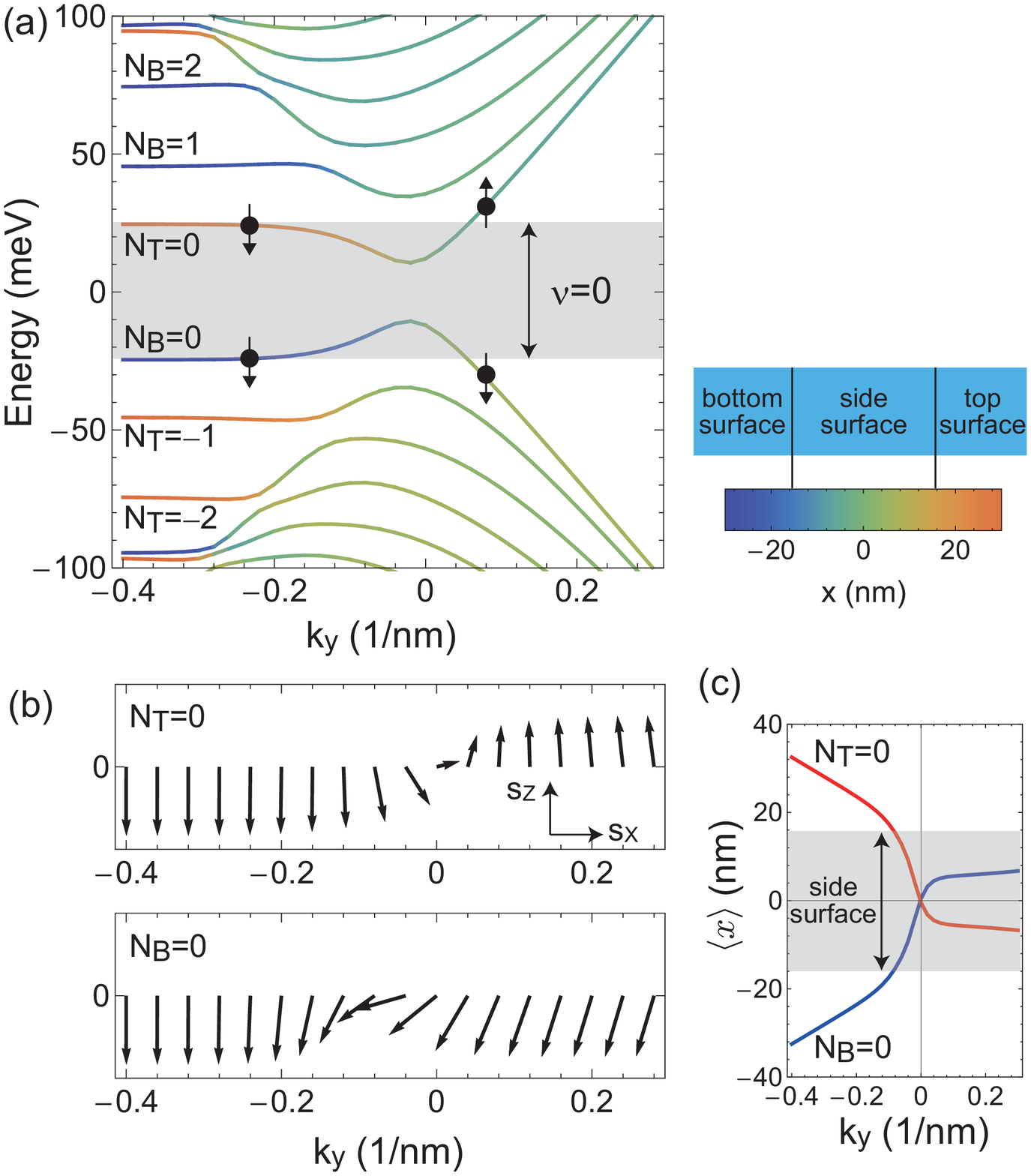}
\end{center}
\caption{
Band structure of the surface Weyl fermions in the 2D model.
(a) The energy levels
of $H$ in the Landau gauge [Eq.~(\ref{eq: Dirac Hamiltonian})]
are plotted
against the momentum $k_y$
for $2d=30$\,nm, $B=15$\,T, $2V_0=50$\,meV.
Colors encode
the expectation value of the position $\braket{x}$;
the Landau levels on the top and bottom surfaces are depicted
in red and blue in the region $k_y<0$,
while dispersive bands in green are edge channels on the side surface.
LLs at the top (bottom) surface are labelled by indices $N_T$ ($N_B$).
(b) 
Spin textures in the $(s_x,s_z)$-plane
and (c) expectation values $\braket{x}$ of the two bands connected
to the $N_T=0$ and $N_B=0$ LLs.
Edge channels ($k_y\gtrsim0$) show the spin-momentum locking,
while the $N=0$ Landau levels on the top and bottom surfaces are 
spin polarized.
}
\label{Fig: band structure}
\end{figure}

\textit{Model ---}
We study quantum Hall states on the surfaces of a TI,
treating the top, bottom, and side surfaces 
as an effective two-dimensional (2D) system
as shown in Fig.~\ref{Fig: surface of TI}.
We will ignore the gapped bulk states.
The top (bottom) surface is mapped onto the region $x>d$ ($x<-d$)
with magnetic fields $+B\hat{\bm{z}}$ $(-B\hat{\bm{z}})$,
while the side surface is mapped onto the region $-d \le x \le d$,
where no magnetic field is applied;
see the lower picture in Fig.~\ref{Fig: surface of TI}.
The effective Hamiltonian for the surface Weyl fermions
is given by
\begin{align}
H= v_F[ -(p_x+ e A_x) \sigma_y + (p_y+ e A_y) \sigma_x] + U \sigma_0,
\label{eq: Dirac Hamiltonian}
\end{align}
where $v_F$ is the Fermi velocity, $\Vec p = (p_x, p_y)$
is the momentum, and
Pauli matrices $\Vec \sigma = (\sigma_x, \sigma_y, \sigma_z)$
act on the spin degrees of freedom, 
and $\sigma_0$ is a $2 \times 2$ unit matrix.
The vector potential $\Vec A = (A_x, A_y)$ and the scalar potential $U$ are
functions of $x$,
\begin{align}
\Vec A(x) &=\biglb(0, - B(x + d)\bigrb), &
U(x) &= - V_0 , &
&(x < -d), \n
\Vec A(x) &=(0, 0), &
U(x) &= 0 , &
&(|x| < d), \\
\Vec A(x) &=\biglb(0, B(x - d)\bigrb), &
U(x) &= V_0 , &
&(x > d). \nonumber
\end{align}
We have introduced the potential difference between the top and
bottom surfaces, $2V_0$.
Since the 2D model is translationally invariant along the $y$ direction
under the Landau gauge,
the momentum $k_y$ is a good quantum number.
The wave functions with wave number $k_y$ have a Gaussian form
in $x$, with the expectation value 
$|\braket{x}| \simeq -k_y \ell^2$ for $k_y<0$ and 
$\braket{x} \simeq 0$ for $k_y>0$,
where $\ell=\sqrt{\hbar/eB}$ is the magnetic length;
see Fig.~\ref{Fig: band structure}(c).

\textit{Edge channels ---}
We show the band structure of the Hamiltonian
[Eq.~(\ref{eq: Dirac Hamiltonian})] in Fig.~\ref{Fig: band structure}(a).
The bands are plotted as functions of $k_y$ for the following parameters:
$2d=30$\,nm, $B=15$\,T, $2V_0=50$\,meV,
and $v_F=5\times 10^5\,\mathrm{m/s}$ \cite{supplement}.
The flat bands at $k_y<0$, shown in red and blue,
are the LLs on the top and bottom surfaces, whose energies are
given by
\begin{align}
E_{N}&=\t{sgn}(N)\sqrt{|N|}\hbar \omega_c + U_0,
\end{align}
with the LL indices $N=N_T$ ($N_B$) and 
the scalar potential $U_0=V_0$ ($-V_0$) for the top (bottom) surface,
and the cyclotron frequency 
$\omega_c=v_F\sqrt{2eB/\hbar}$.
Shown in green/yellow at $k_y \gtrsim 0$ are the
energy bands of Weyl fermions on the side surface,
which are separated by the finite-thickness effects.
The band structure in Fig.~\ref{Fig: band structure}(a)
can be naturally obtained by smoothly connecting
LLs on the top and bottom surfaces to 
the discrete energy bands of Weyl fermions of the side surface.
The $N_{T/B}=0$ LLs on the top and bottom surfaces are separated
by the energy difference $2V_0$.
Thus, the $\nu=0$ QHE is realized when the Fermi energy $E_F$ is
in the energy window $-V_0<E_F<V_0$,
because LLs are filled up to $N_T=-1$ and $N_B=0$ LLs at top and bottom
surfaces, respectively, and hence $\nu=n_T+n_B+1=0$ with $n_T=-1$ and $n_B=0$.
In this case there is no chiral edge mode,
as expected from the bulk-edge correspondence.
However, transport through edge channels
is still possible in the $\nu=0$ QHE,
under the conditions we discuss below.

Three parameters control the existence of edge channels 
in the $\nu=0$ QHE:
the energy difference of the Weyl points of the top
and bottom surfaces $2V_0$,
the cyclotron energy $\hbar \omega_c$,
and the energy gap at $k_y=0$ of the Weyl fermions on the side surface
$\Delta_{\t{side}}\simeq \hbar v_F/2d$.
The edge channels in the $\nu=0$ QHE can exist when
\begin{align}
\Delta_\mathrm{side} < 2V_0 \le \hbar \omega_c. 
\label{eq: condition for edge channels}
\end{align}
The left inequality implies that the two bands from the $N_{T/B}=0$
LLs turn into helical edge channels on the side surface with
energies in the $\nu=0$ quantum Hall regime ($|E|<V_0$),
whereas the right inequality assures that the $N_{T/B}=0$ LLs
are located in between the $N_T=-1$ and $N_B=+1$ LLs.
The above conditions are satisfied by the parameters used in
Fig.~\ref{Fig: band structure}(a), which shows edge modes
with a gap at $k_y=0$.
Furthermore, we find that, for the parameters
in the experiment of (Bi$_{1-x}$Sb$_x)_2$Te$_3$ in Ref.~\cite{Yoshimi},
the above inequalities are satisfied with
$2V_0 = 70 \,\t{meV}$, 
$\hbar\omega_c = 70 \,\t{meV}$, and
$\Delta_\mathrm{side} \simeq 40 \,\t{meV}$.

In Fig.~\ref{Fig: band structure}(b)
we show the expectation values of 
$
(\braket{s_x}, \braket{s_z})
$
for the Bloch wave functions of
the valence top band and the conduction bottom band,
connected to the $N_B=0$ and $N_T=0$ LLs, respectively;
we find $\braket{s_y}=0$ for all bands.
Note that, since we have rotated the side and bottom surfaces
around the $y$ axis
by $90^\circ$ and $180^\circ$, respectively, 
in our 2D model of TI surfaces (Fig.~\ref{Fig: surface of TI}),
the spins are also rotated accordingly.
Thus the real spin operators $\Vec{s}$ are written in terms of 
the Pauli matrices in the Hamiltonian 
[Eq.~(\ref{eq: Dirac Hamiltonian})] as
\begin{align}
\frac{\Vec{s}}{\hbar/2} &=
\begin{cases}
(\sigma_x, \sigma_y, \sigma_z), & (x>d),\\
(-\sigma_z, \sigma_y, \sigma_x), & (-d<x<d),\\
(-\sigma_x, \sigma_y, -\sigma_z), & (x<-d).
\end{cases}
\end{align}
The spins of the $N_{T/B}=0$ LLs on the top and bottom surfaces
are polarized,
$(\braket{s_x}, \braket{s_z})=(0,-\hbar/2)$ for $k_y<0$.
As $k_y$ increases, the spin configuration crosses over to
a helical spin structure of the edge channels on the side surface
for $k_y\gtrsim0$, where
the spins of the two bands are polarized to the opposite directions
because of the spin-momentum locking.
This feature is similar to helical edge states of quantum spin Hall insulators,
except for the presence of the gap at $k_y=0$.
We note that, when $V_0$ is negative, 
all the discussion above applies by replacing $V_0$
by $-V_0=|V_0|$,
except that $\braket{s_x}$ becomes $-\braket{s_x}$ in the spin texture.
This is because the reflection $z \to -z$ changes the sign of $V_0$
and also $(s_x, s_y) \to (-s_x, -s_y)$.

\begin{figure}[tb]
\begin{center}
\includegraphics[width=\linewidth]{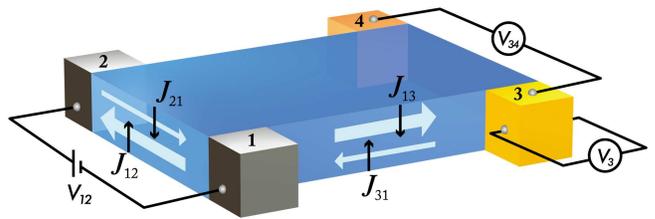}
\end{center}
\caption{
A setup for the detection of charge and spin transport through
edge channels.
Applying a voltage between the electrodes 1 and 2 in the left
induces charge and spin transport through the edge 
channels in the $\nu=0$ quantum Hall state.
The nonlocal charge transport is measured by the voltage
$V_{34}$ between the electrodes 3 and 4.
The spin current is detected by the voltage
$V_3$ between the two sides of the Pt electrode 3
via the inverse spin Hall effect.
}
\label{Fig: spin detection}
\end{figure}

\textit{Charge and Spin transport by edge channels ---}
The edge channels on the side surface are responsible for
the charge and spin transport,
when the Fermi energy is in the range
$\Delta_\mathrm{side}<|E_F|<V_0$.
(Unlike in the quantum Hall insulators, however,
the charge and spin conductances are not quantized
because backscattering is allowed in these edge channels.)
Here we propose an experimental setup consisting of
a TI thin film with four electrodes to observe
the charge/spin transport through edge channels;
see Fig.~\ref{Fig: spin detection}.
In the following we will discuss three characteristic transport phenomena:
charge transport, spin transport, and spin conversion.

First, the edge channels should lead to nonlocal transport of charge.
We predict that, when a voltage is applied
and electric current flows between the electrodes 1 and 2,
a finite voltage $V_{34}$ between the electrodes
3 and 4 should be observed.
Let us assume
$V_0<E_F<\Delta_\mathrm{side}$.
The applied voltage induces net flow of up-spin electrons to the $+y$
direction and that of down-spin electrons to the $-y$ direction,
according to Fig.~\ref{Fig: band structure}.
This results in the flow of electrons to the directions indicated
by the thick arrows $J_{12}$ and $J_{13}$ in Fig.~\ref{Fig: spin detection}
(note that the electric current flows to the opposite directions),
while the bulk transport is vanishing.
The thin arrows represent the residual flow of electrons which were not absorbed by electrodes.
The current $J_{13}$ flows further along the right and rear side surfaces
before reaching the electrode 2, yielding a finite voltage $V_{34}$ between
the electrodes 3 and 4.
The nonlocal transport in the $\nu=0$ quantum Hall regime
has been actually observed in experiments of (Bi$_{1-x}$Sb$_x$)$_2$Te$_3$
films \cite{PrivateCommun}.

\begin{figure}[tb]
\begin{center}
\includegraphics[width=\linewidth]{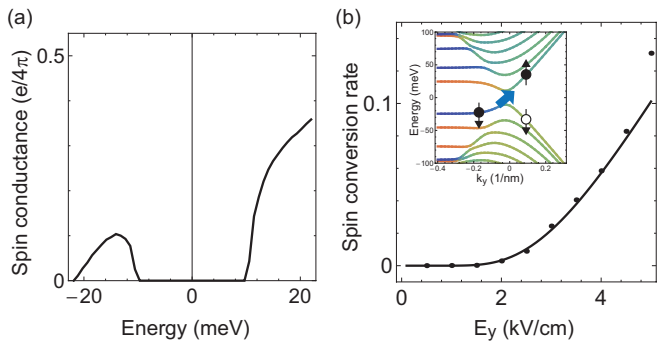}
\end{center}
\caption{
(a) Spin conductance as a function of the Fermi energy.
Spin conductance is finite in the $\nu=0$ quantum Hall state
when the Fermi level is in the edge channels.
(b) Spin conversion rate plotted against electric fields $E_y$.
Dots represent the spin conversion rate
obtained from numerical solutions of the time-dependent Schr\"odinger equation,
while the solid curve is given by the Zener tunneling formula
[Eq.~(\ref{eq: Landau-Zener formula})].
Results in both panels are obtained for the same parameters
used in Fig.~\ref{Fig: band structure}(a).
}
\label{Fig: spin conductance and conversion}
\end{figure}

Second, the edge channels support spin transport
in addition to the charge transport.
As we have discussed above,
the applied voltage $V_{12}$ produces the circulating
spin polarized electronic current through the edge channels,
as shown in Fig.~\ref{Fig: spin detection}.
We predict that the nonvanishing spin current should yield
a finite voltage $V_3$ between the two sides of the electrode 3
if it is made of a material with strong spin-orbit coupling such as Pt.
The spin current $J_{13}$ diffusing into the electrode 3 produces
a flow of down-spin electrons moving perpendicular to the side surface.
This spin current will generate the voltage
$V_3$ through the inverse spin Hall effect (ISHE) \cite{Saitoh}
 in the electrode 3.
The voltage $V_3$ is determined by
the spin conductance of the edge channels,
the diffusion rate across the interface of the electrode 3,
and the efficiency of the ISHE.
The (linear) spin conductance $G_{s_z}$ is calculated
in the ballistic regime as
\begin{align}
G_{s_z}=\sum_i\int \langle s_z \rangle_{i,k_y} |v_{i,k_y}|
\left(-\frac{e}{2}\right)\!\frac{df(\epsilon_{i,k_y})}{d\epsilon}dk_y,
\label{G_s}
\end{align}
where $f(\epsilon)$ is the Fermi distribution function, and
$\langle s_z \rangle_{i,k_y}$ is the expectation value of $s_z$ 
for the Bloch wave function $\psi_{i,k_y}$ 
of the $i$th band with the momentum $k_y$
and the energy $\epsilon_{i,k_y}$.
The group velocity $v_{i,k_y}$ along the $y$ direction is given by
$v_{i,k_y}=\hbar^{-1} d \epsilon_{i,k_y}/dk_y$.
In deriving Eq.\ (\ref{G_s}) we have assumed that
the applied voltage $V$ shifts the Fermi energy of the right-going
(left-going) electrons by $V/2$ ($-V/2$) in the ballistic regime.
In Fig.~\ref{Fig: spin conductance and conversion}(a)
we show the spin conductance at zero temperature
which is calculated as a function of $E_F$ with the parameters used in
Fig.~\ref{Fig: band structure}(a).
The spin conductance $G_{s_z}$ is non-vanishing when the Fermi energy
lies in the edge channels.
The asymmetry of the spin conductance in $E_F$ reflects the
different spin configurations of the two bands connected to
the $N_{T/B}=0$ LLs shown in Fig.~\ref{Fig: band structure}(a).

Finally, we discuss spin conversion due to interband transitions
that occur at the side surface in the non-equilibrium regime.
We assume that the Fermi energy is within the gap $\Delta_{\t{side}}$.
When an electric field is applied, by external gate electrodes,
to the sample in the direction from the electrode 2 to the electrode 1,
electrons in the $N_B=0$ band are driven by the electric field
and undergo Zener tunneling into the $N_T=0$ band
across the energy gap around $k_y = 0$
as shown in the inset of 
Fig.~\ref{Fig: spin conductance and conversion}(b),
where $k_y$ is along the direction from the electrode 1 to the
electrode 2.
In this process, a pair of a spin-up electron in the $N_T=0$ band 
and a spin-down hole in the $N_B=0$ band is created.
Thus the Zener tunneling in the edge channels 
gives rise to spin conversion and spin accumulation at the side surface
between the electrodes 1 and 2.
In order to quantify the spin conversion,
we study the time-evolution of a wave packet
driven by the electric field $E_y$ along the $y$ direction
for the time period $t_f=\hbar(k_y^f-k_y^i)/eE_y$,
from the initial state
$\psi(t=0)=\psi_{N_B=0,k_y^i}$ in the $N_B=0$ band
with the momentum $k_y^i = -0.2~\t{nm}^{-1}$
to the final state $\psi(t=t_f)$ 
with the momentum $k_y^f = 0.2~\t{nm}^{-1}$.
To this end, we numerically solve the time-dependent Schr\"odinger equation,
\begin{align}
i\hbar \partial_t \psi(t) &=(H + v_F e E_y t  \sigma_x) \psi(t),
\end{align}
and compute the spin conversion rate $P(E_y)$
\begin{align}
P(E_y)&= \braket{s_z}_{t=t_f} - \braket{s_z}_{N_B=0,k_y^f},
\end{align}
where $\braket{s_z}_{t=t_f}$ is
the expectation value of $s_z$ in the final state $\psi(t_f)$.
The obtained spin conversion rate in
Fig.~\ref{Fig: spin conductance and conversion}(b)
shows a nonlinear response to $E_y$,
which is consistent with
the Zener tunneling formula,
\begin{align}
P(E_y)& \propto \exp\!\left(-\frac{2\pi}{\hbar}
\frac{\Delta_{\t{side}}^2}{2 v_F eE_y}\right),
\label{eq: Landau-Zener formula}
\end{align}
for the two-band system of 
the $N_T=0$ and $N_B=0$ bands
with the energy gap $\Delta_{\t{side}} \simeq v_F\hbar/2d$.

\textit{Summary ---}
We have studied charge and spin transport
in the $\nu=0$ quantum Hall system of surface Weyl fermions of 3D TIs.
The charge and spin currents are carried by the edge channels
formed on the side surfaces of TIs.
A unique feature of the QHE in TIs is that the top
and bottom surfaces, where LLs are formed, are
connected by side surfaces, where Weyl fermions form
edge channels,
as contrasted to conventional quantum Hall bilayers and
the $\nu=0$ QHE in graphene.
In the proposed spin transport experiment,
the voltage $V_3$ is expected to be on the order of $10$\,nV 
as measured in an experiment of ISHE \cite{Omori}.
The spin conversion rate is typically $0.1 \sim 1$\,\%
under the electric field $E_y \simeq 1$\,kV/cm,
when electric current of a few mA flows in a 1 mm sample \cite{Yoshimi}.
We expect that the spin accumulation can be detected by Kerr rotation of
the order of $10\,\mu$rad at the edge of TI thin films \cite{Kato}.

The authors are grateful for insightful discussions with M. Kawasaki and Y.
Tokura. This work was supported by Grant-in-Aids for
Scientific Research (No.~24224009 and No.\ 24540338) from the Ministry 
of Education, Culture, Sports, Science and
Technology (MEXT) of Japan
and from Japan Society for the Promotion of Science.


\clearpage
\widetext
\begin{center}
\textbf{\large Supplemental material for 
``Charge and spin transport in edge channels of 
a $\nu=0$ quantum Hall system on the
surface of topological insulators''}
\end{center}
\setcounter{equation}{0}
\setcounter{figure}{0}
\setcounter{table}{0}
\makeatletter
\renewcommand{\theequation}{S\arabic{equation}}
\renewcommand{\thefigure}{S\arabic{figure}}
\renewcommand{\bibnumfmt}[1]{[S#1]}
\renewcommand{\citenumfont}[1]{S#1}

\subsection{LLs in top and bottom surfaces}
Far from the side surface,
we have well-defined Landau levels (LLs) at the top and bottom surfaces.
In this region,
the Hamiltonian [Eq.~(1)] reduces to
\begin{align}
H_{k_y}=v_F[
-p_x \sigma_y +
(\hbar k_y \pm e B x) \sigma_x
]
\pm V_0 \sigma_0,
\end{align}
where $k_y$ is the momentum along the $y$ direction and we choose the sign $\pm$ for the top and bottom surfaces.
We note that we replaced the gauge potential with $A_y=\pm Bx$ 
for simplicity.
We assume $B>0$.
The energy and wavefunction of LLs for the above Hamiltonian 
are written as follows.
For the top surface,
the LLs are written as
\begin{align}
E_{N_T,k_y}&=\t{sgn}(N_T)\sqrt{|N_T|}\hbar\omega_c + V_0, \n
\psi_{N_T,k_y}(x)&=
\begin{pmatrix}
\t{sgn}(N_T) \phi_{|N_T|-1} \left(\frac{x-x_0(k_y)}{\ell} \right) \\
\phi_{|N_T|} \left(\frac{x-x_0(k_y)}{\ell} \right) \\
\end{pmatrix}, \n
x_0(k_y)&=-\ell^2 k_y,
\end{align}
with
\begin{align}
\t{sgn}(n)=
\begin{cases}
+1, &\quad (n>0) \\
0,  &\quad (n=0)  \\
-1, &\quad (n<0) \\
\end{cases}
\end{align}
where $N_T$ is the Landau index at the top surface, 
$\phi_n$ is the wavefunction of conventional Landau levels with $\phi_{-1}=0$,
the cyclotron energy $\hbar\omega_c$ and the magnetic length $\ell$ are
\begin{align}
\hbar\omega_c &=v_F \sqrt{2\hbar eB}, &
\ell &=\sqrt{\frac{\hbar}{eB}}.
\end{align}
For the bottom surface, the LLs are written as
\begin{align}
E_{N_B,k_y}&=\t{sgn}(N_B)\sqrt{|N_B|}\hbar\omega_c - V_0 , \n
\psi_{N_B,k_y}(x)&=
\begin{pmatrix}
\phi_{|N_B|} \left(\frac{x-x_0(k_y)}{\ell} \right) \\
\t{sgn}(N_B)\phi_{|N_B|-1} \left(\frac{x-x_0(k_y)}{\ell} \right) \\
\end{pmatrix}, \n
x_0(k_y)&=\ell^2 k_y.
\end{align}
where $N_B$ is the Landau index at the bottom surface.

\subsection{Band structure with parameters in an experiment of TI thin films}
\begin{figure}[tb]
\begin{center}
\includegraphics[width=0.5\linewidth]{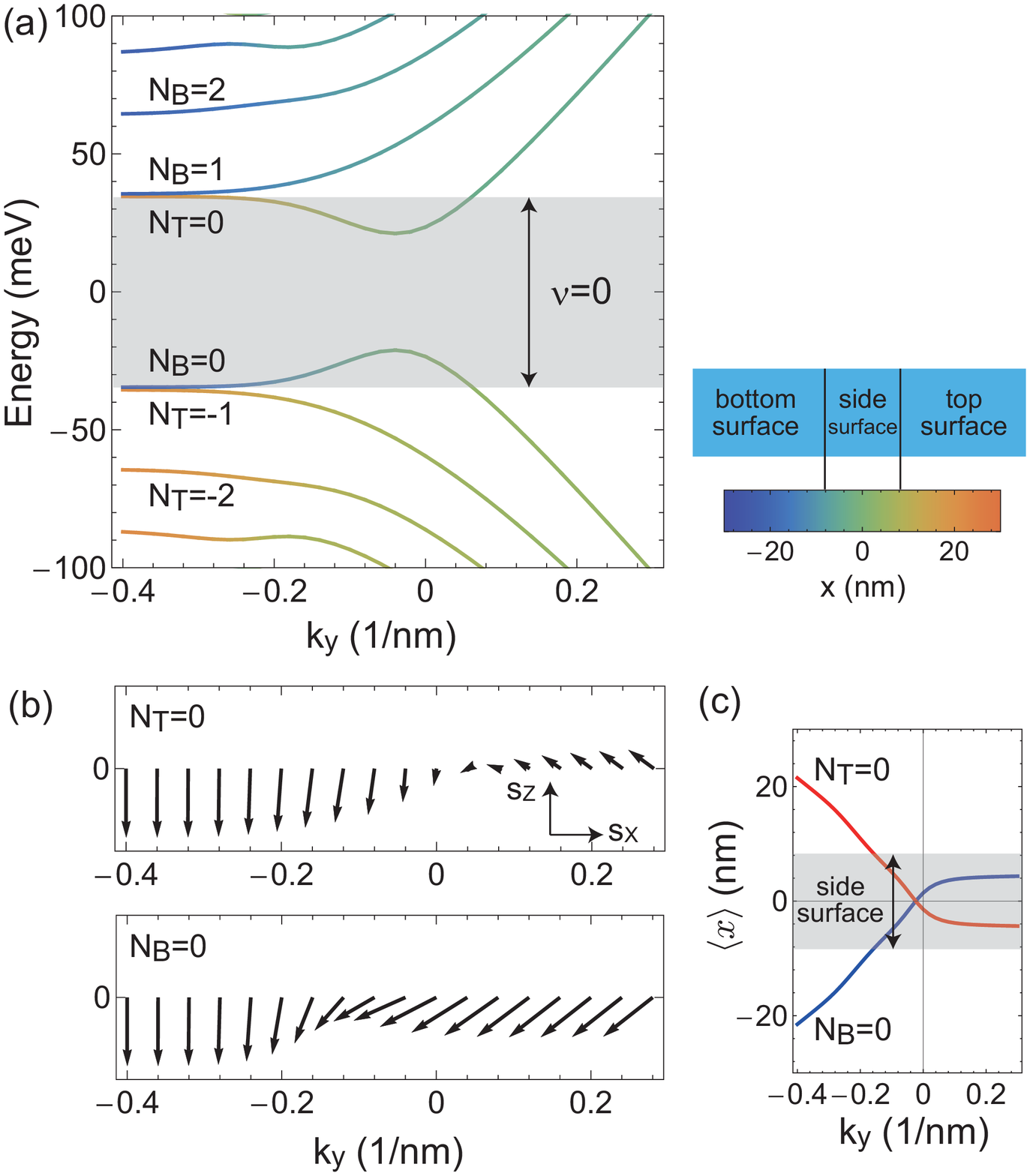}
\end{center}
\caption{
Band structure of the surface Weyl fermions in the 2D model.
(a) The energy levels
of $H$ in Eq.~(1)
are plotted
against the momentum $k_y$
for $2d=8$\,nm, $B=15$\,T, $2V_0=70$\,meV.
Colors encode
the expectation value of the position $\braket{x}$;
the Landau levels on the top and bottom surfaces are depicted
in red and blue in the region $k_y<0$,
while linear bands in green are edge channels on the side surface.
LLs at the top (bottom) surface are labelled by indices $N_T$ ($N_B$).
(b) 
Spin textures in the $(s_x,s_z)$-plane
and (c) expectation values $\braket{x}$ of the two bands connected
to the $N_T=0$ and $N_B=0$ LLs.
Edge channels ($k_y\gtrsim0$) show the spin-momentum locking,
while the $N=0$ Landau levels on the top and bottom surfaces are 
spin polarized.
}
\label{Fig: band structure app}
\end{figure}

We show the band structure of the Hamiltonian [Eq.~(1)] using the parameters in experiments by Yoshimi et al. \cite{Yoshimi-SM}:
$2d=8$\,nm, $B=15$\,T, $2V_0=70$\,meV,
and $v_F=5\times 10^5\,\mathrm{m/s}$.
The band structure is plotted as a function of $k_y$ in Fig.~\ref{Fig: band structure app}(a).
We find that $N_B=0$ and $N_T=-1$ LLs
and $N_B=1$ and $N_T=0$ LLs are almost degenerate.
Edge channels appear in the $\nu=0$ quantum Hall system (QHS) 
in the energy window $-V_0<E_F<V_0$.
The condition for the presence of edge channels [Eq.~(4)] are satisfied with
$\Delta_{\t{side}}\simeq 40 \t{meV} < 2V_0$.
This indicates that nonlocal transport observed in Ref.~\cite{PrivateCommun-SM} originates from the edge channels in the $\nu=0$ QHS.
The spin expectations values in Fig.~\ref{Fig: band structure app}(b) show that
edge channels are spin-momentum locked 
and the $N=0$ LLs are spin-polarized.
Thus observations of the spin transport phenomena proposed in this paper are 
experimentally feasible in 
available topological thin films with the experimental setups in Fig.~3.

Estimates of the parameters controlling presence or absence of the edge channels 
are given for $v_F=5\times 10^5\,\mathrm{m/s}$ as follows.
The cyclotron energy $\hbar \omega_c$ is given by
\begin{align}
\hbar \omega_c 
\simeq 18 \sqrt{B(\mathrm{T})} \, \t{meV}.
\end{align}
The gap of the Weyl fermions at the side surface due to the finite size effect can be roughly estimated as
\begin{align}
\Delta_\mathrm{side} \simeq
 \frac{\hbar v_F}{2d} \simeq \frac{320}{2d(\mathrm{nm})} \t{meV}.
\end{align}

\end{document}